\documentstyle[sprocl,amstex,epsfig]{article}

\def\l{\label}
\def\n{\nonumber\\}
\def\g{\gamma}
\def\D{\Delta}
\def\a{\alpha_s}

\begin{document}

\title{Unified approach to hard diffraction
}

\author{R.Peschanski}

\address{Service de Physique Th\'eorique, CEA, CE-Saclay\\
F-91191 Gif-sur-Yvette Cedex, France}


\maketitle\abstracts{Using a combination of S-Matrix and perturbative QCD 
properties in the small 
$x_{Bj}$ regime, 
we propose a formulation of hard diffraction unifying the partonic 
(Ingelman-Schlein) Pomeron, 
Soft 
Colour Interaction and QCD dipole descriptions.}

\section{Introduction}

 In the present paper, we  focus on three existing  different theoretical 
approaches 
of hard diffraction, for which we propose a new, unifying, formulation. The 
first  one we will 
refer to is the ``partonic Pomeron'' approach\cite 
{in85}.  The 
hard photon  is here supposed to probe the parton distributions of the Pomeron 
Regge pole considered as a hadronic particle. A second approach\cite{in96} is 
the Soft Colour Interaction one , 
where hard diffraction is described by a two-step process: during a relatively 
short 
``interaction 
time'', the probe initiates a typical hard deep-inelastic interaction. Then, at 
large 
times/distances, a soft colour 
interaction 
is assumed which will decide of  the separation between diffractive and 
non-diffractive events according to a simple probabilistic rule.
 A third approach is based on the small $x_{Bj}$ regime of perturbative QCD, 
where the 
resummation of leading $\log 1/x_{Bj}$ leads to the ``QCD dipole'' 
approach for hard diffraction\cite {bi96}.

We show that the three aproaches may find a common 
formulation. Our main results are the following:

 {\bf 1)} The ``effective'' parameters of 
the partonic Pomeron are determined from  
leading 
log perturbative  QCD 
resummation. They are found 
to depend not only on  $Q^2$ but also on the ratio
$\eta = 
 \left(Y-y\right) /y,$ where Y (resp. y) are the total (resp. gap) rapidity 
interval.
 
{\bf 2)} The ``effective'' parameters of 
the hard process before modification by the Soft Colour Interaction are 
determined. Differences with the original model  avoid the ``Low's theorem'' 
paradox\cite {dok}.
 
 {\bf 3)} Using S-Matrix properties of triple-Regge contributions, a relation is 
found 
between  discontinuities of a $3 \to 3$ amplitude and the three  approaches to 
hard 
diffraction. In particular, a theoretically founded  formulation of a (modified) 
Soft Colour Interaction approach is 
proposed as a 
specific 
double discontinuity the $3 \to 3$ forward amplitude.

\section{From  QCD dipoles to partonic Pomeron}

Our  starting point\cite{na01}   is the formula for the
 structure function 
(inelastic component) 
 for longitudinal and 
transverse photon:

\begin{eqnarray}
F_{T,L}^{Diff}(Q^2,Y,y)\sim
\int_{c-i\infty}^{c+i\infty}\frac 
{d\g_1}{2i\pi}\frac {d\g_2}{2i\pi}\frac {d\g}{2i\pi}\ \delta 
(1\!-\!\g_1\!-\!\g_2\!-\!\g)\ 
 \n
\times \left(\frac Q{Q_0}\right)^{2\g}\ \exp \left\{y [\D(\g_1)+\D(\g_2)] + 
(Y\!-\!y)\ 
[\D(\g)]\right\}\ ,
\l{biczy}
\end{eqnarray}
where
\begin{equation}
\D(\g) \equiv \frac {\a 
N_c}{\pi}\left\{2\psi(1)-2\psi(\g)-2\psi(1-\g)\right\} \sim \D + \frac 
{\D^{''}}2 \left(\frac 12\!-\!\g \right)^2
\l{chi}
\end{equation}
 is the BFKL evolution kernel (used in a gaussian approximation), $Q_0$ a 
non-perturbative scale associated with the proton. After a triple saddle-point 
approximation (using the gaussian approximation) justified by the 
 large rapidity gap characteristic of diffraction, one may write $
F_{T,L}^{Diff} \ \sim
e^{2y\Delta } \ \sigma^{tot}_{\gamma^*-P}\ ,$
with 
\begin{equation}
\sigma^{tot}_{\gamma^*-P} \sim \sqrt {\frac {2}
{1\!+\!2\eta}}
\  {\exp \left\{ (Y\!-\!y) {\bf \epsilon}   _s \right\} }\left(\frac 
Q{Q_0}\right)^{2{\bf \g_s}}
\!\exp \left( -\frac  {2\eta}{1\!+\!2\eta} \frac {2\log^2\left(\frac 
Q{Q_0}\right)}{\D 
^{''} (Y\!\!-\!y)}\right) \ .
\l{biczy3} 
\l{QCDRegge}
\end{equation}
 
 Here, the gaussian saddle-points determine both  the new ``effective 
dimension'' and 
``effective 
intercept'':
\begin{equation}
{\bf \g_s}= \frac {\eta}{1+2\eta}\ ;\ 
{\bf \epsilon}_s  =\D + \frac {\D^{''}}{8 (1+2\eta)}
\l{epsilon}
\end{equation}
of a ``BFKL-like'' expression, depending on the variable  
$\eta = \frac {Y\!-\!y}{y}$.

$\sigma^{tot}_{\gamma^*-P}$ thus defines an effective ``partonic Pomeron'' 
derived from the QCD dipole formalism. 

\section{From dipoles to Soft Colour Interaction}

In Soft Color Interactions models, at least in their formally simpler 
formulations, one expects the following relation between the total structure 
function and the overall contribution of hard diffraction  at 
fixed value 
of $x_{Bj}:$

\begin{equation}
F_{T,L}^{Diff/tot}\!\! \equiv \!\int_{x_{Bj}}^{x_{gap}} dx_P \ F_{T,L}^{Diff} = 
{\bf \frac 1{N_c^2}}\ 
F_{T,L}^{hard/tot}
\l{biczy4}
\end{equation}
where $\log\frac 1{x_{gap}}$ is the 
minimal value retained for 
the rapidity 
gap. {\it A priori}, $F_{T,L}^{hard/tot}$ is supposed not to be modified by the 
soft color interaction. However a contradiction with ``Low's theorem'' has been 
quoted\cite{dok}. Indeed, a fully soft color interaction can be proven to give 
no effect on physical obervables. 

In fact, when inserting the QCD dipole expression (\ref{biczy}) in relation 
(\ref{biczy4}), one finds\cite{na01}
\begin{equation}
F_{L,T}^{hard/tot} \sim \left(\frac 
Q{Q_0}\right)^{2\g_{hard}}
\frac {\exp \left(Y\D (\g_{hard})\right)}{\sqrt {2\pi\D ^{''} \ Y}}  \ ,
\l{tot1}
\end{equation}

which is very similar with the  BFKL expression, except that 
\begin{equation}
\g_{BFKL} \sim \frac 12 \left( 1-  4\frac 
{\log\left(\frac Q{Q_0}\right)}{\D ^{''} \ Y}
\right)\ \ne \g_{hard} = cst \sim 0.175.
\l{value}
\end{equation}

Thus, in some sense, the Soft Color relations are compatible with QCD dipole 
calculations, but the underlying hard interaction is somewhat different than the 
assumed mechanism in SCI models. Hence the soft color rearrangement is not 
simply a  ``neutralization'' of color at long distances which would be forbidden 
by virtue of ``Low's theorem''. Apart this modification, formula (\ref{biczy4}) 
fixes the relative normalization between diffractive and total structure 
functions.
.
\section{S-Matrix interpretation}

Following old results of S-Matrix theory in the Regge domain\cite {mu}, and as 
sketched in 
the figure , one may 
consider three types of discontinuities of a $3 \to 3$ amplitude representing 
hard diffraction. A single discontinuity over the diffractive $mass^2$ variable, 
a double discontinuity  taking into account also the 
analytic discontinuity  in the subenergy of one of the incident
 Pomeron exchanges 
and a triple discontinuity   with the discontinuity including
the two 
Pomeron 
exchanges.

\begin{figure}[t]
\begin{center}
\mbox{\epsfig{figure=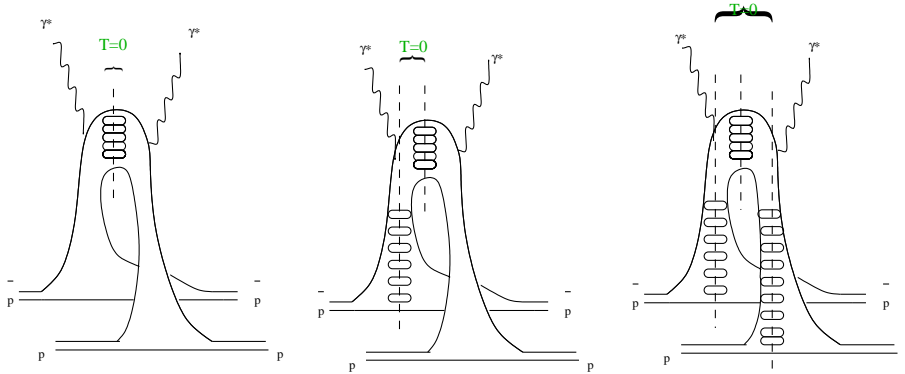,height=5.0cm}}
\vspace{-0.5cm}
\caption{
``Time-dependent'' picture of the 
$A(3 \to 3)$ discontinuities.
Upper graph: Description of $Disc_1 A(3 \to 3)$ (partonic Pomeron 
approach); Middle 
graph: Description of $Disc_2 A(3 \to 3)$ (candidate for the Soft Colour 
Interaction approach); Lower graph: 
Description of $Disc_3 A(3 \to 3)$ (QCD dipole approach).}
\vspace{0.5cm}
\label{fig:fig1}
\end{center}
\end{figure}

As qualitatively shown in the figure, one can associate the three discontinuity 
formulae with, respectively, the partonic Pomeron, Soft Color Interaction and 
QCD dipole approaches. This is schematically depicted in 
the ``time-dependent'' representation of the figure, where the soft incident 
Pomerons are 
represented 
as living during a ``long'' time, while the partonic (or QCD dipole) process are
initiated by  hard interactions. In fact, one can prove\cite{mu} the identity of 
these discontinuities, and thus legitimate the connection realized in the 
previously mentionned calculations. 
\section*{Acknowledgments}

This work was done in  collaboration with H. Navelet.

\end{document}